# Effect of magnetic field and temperature on the ferroelectric loop in MnWO$_4$

Bohdan Kundys,[*] Charles Simon, and Christine Martin

*Laboratoire CRISMAT, CNRS UMR 6508, ENSICAEN, 6 Boulevard du Maréchal Juin, 14050 Caen Cedex, France*

The ferroelectric properties of MnWO$_4$ single crystal have been investigated. Despite a relatively low remanent polarization, we show that the sample is ferroelectric. The shape of the ferroelectric loop of MnWO$_4$ strongly depends on magnetic field and temperature. While its dependence does not directly correlate with the magnetocapacitance effect before the paraelectric transition, the effect of magnetic field on the ferroelectric polarization loop supports magnetoelectric coupling.

The mutually exclusive nature of magnetism and electric polarization phenomena in most solids[1,2] has recently attracted attention in the scientific community. This is essentially due to both the basic physics challenges posed and the possible magnetoelectric (ME) applications for memory storage and electric field-controlled magnetic sensors. The idea of having the two order parameters (magnetic and electric) at the same temperature and magnetically controlled electrical polarization (or vice versa) has stimulated a vast research of new materials,[3–8] as well as reinvestigation of previously known compounds.[9] It becomes evident that many canted antiferromagnets may develop electric polarization as a result of the overlap of the electronic wave functions and as a result of the spin orbit interaction.[10] Among materials in which magnetoelectric effects have been recently reported, MnWO$_4$ is a particularly interesting material as the electric polarization in a single crystal may be switched from the $b$ to $a$ direction when a strong magnetic field is applied.[11–14] A similar phenomenon has been observed in rare-earth manganites.[15,16] The antiferromagnetic (AF) phase transitions of MnWO$_4$ were already studied a long time ago.[17,18] With decreasing temperature, MnWO$_4$ undergoes a collinear antiferromagnetic state (AF1 phase) at $T_N \approx 13.5$ K with further transformation to the noncollinear incommensurate antiferromagnetic phase (AF2 phase) at 12.7 K, and finally to collinear incommensurate antiferromagnetic phase (AF3 phase) at 7.6 K. Among the three antiferromagnetic states, only the noncollinear one (AF2 phase) appears to develop an electric polarization that can be explained within the framework of the phenomenological[10,19–21] and microscopic models.[22] Polarity alone, however, does not guarantee ferroelectricity that is sometimes difficult to experimentally demonstrate.[9,23] For example, the reversibility of the electric dipoles could require electric fields larger than the breakdown field, or it might be due to asymmetric irreversible arrangements of the atoms. In this Brief Report, the electric field-induced dipole reversibility (ferroelectricity) of MnWO$_4$ is shown, along with the measures of its dependence on magnetic field and temperature. dc ferroelectric measurements were performed in a Physical Property Measurement System (PPMS) Quantum Design cryostat by using a Keithley 6517A electrometer. The used technique is our adaptation of the already known method,[24] wherein programming technology has been applied to the Keithley 6517A electrometer and PPMS to provide the possibility of temperature- and magnetic field-dependent studies of the ferroelectric polarization loops. Its quasistatic (dc) operational nature allows ferroelectric loops to be observed with a small electric fatigue risk at ultralow frequency measuring signals. A 0.37 mm sized along the $b$ axis single crystal of MnWO$_4$, which is grown by the floating zone method, has been cut for ferroelectric loop measurements. The magnetic and electric fields were applied parallel to the direction of the crystalline $b$ axis. The electrical contacts with the sample were made by using a conductive silver paint. Figure 1 presents the ferroelectric hysteresis loops $[P(E)]$ as a result of current-voltage $[I(E)]$ integration with respect to time at dif-ferent temperatures recorded after a zero electric and magnetic field cooling procedure. The remanent polarization of about 39 $\mu C/m^2$ at 10 K agrees well with the reported forced polarization in this material.[11,12]

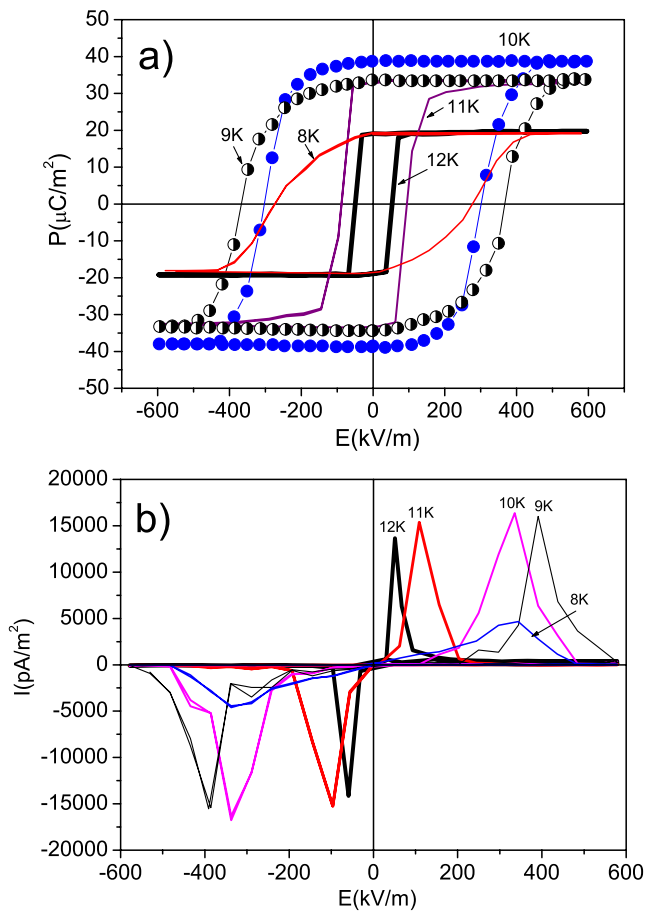

FIG. 1. (Color online) (a) Ferroelectric loops obtained as a result of current integration at different temperatures. (b) Corresponding voltage-current characteristics taken at different temperatures.

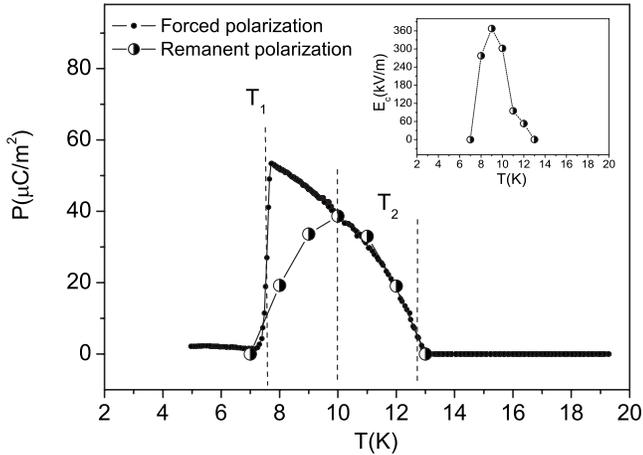

FIG. 2. Remanent polarization extracted from ferroelectric $P(E)$ loops and the forced polarization recorded at heating with the previous electric (520 kV/m) cooling procedure. The inset shows the ferroelectric coercive force as a function of temperature.

Remanent polarization (Fig. 2) and ferroelectric coercive force (inset of Fig. 2) extracted from ferroelectric loops go through a maximum and decrease to zero for temperatures close to the magnetic transitions (i.e., 7.6 and 12.7 K). We have also measured the forced polarization upon heating [electric field (520 kV/m) cooling procedure] (Fig. 2). Near 7.6 K, the forced polarization more rapidly increases than the remanent one (Fig. 2) and the maximum is reached at 8 and 10 K for the forced and remanent polarizations, respectively. This experimental result indicates that the ability to switch the ferroelectric polarization with electric field more quickly vanishes than the forced electric polarization in the sample in this temperature region.

The effect of the external magnetic field applied along the $b$ axis on ferroelectric switching processes [$I(E)$ and $P(E)$ loops] at 10 K is shown in Fig. 3. The ferroelectric coercive force and the remanent polarization decreased upon external magnetic field application, and the ferroelectric loop is no more observed at 12 T. The magnetic field dependence of the dielectric permittivity at 10 K and the magnetic field dependence of the ferroelectric coercive force (along the $b$ axis) are shown in Fig. 4. The position of the peak in the dielectric permittivity shows no hysteresis with respect to the magnetic field and agrees well with the magnetic field dependence of the both the ferroelectric coercive force and the remanent polarization (not shown). While the peak in the magnetic field dependence of dielectric permittivity is very narrow along the $a$ axis,[11] a rather broad anomaly in the dielectric permittivity is observed along the $b$ crystallographic direction (Fig. 4). It is also worth noting that practically no magnetodielectric effect is seen in magnetic fields up to 9 T (Fig. 4), while the magnetic field-induced change in the shape of a ferroelectric loop (values of the remanent polarization and of the ferroelectric coercive force) indicates a magnetoelectric coupling in this magnetic field range [see Fig. 3(a)]. This behavior is in agreement with the identical slope of ferroelectric loops near zero electric field for magnetic fields less than 9 T [Fig. 3(a)]. These results, therefore, imply that magnetoelectric interactions are present without noticeable magnetodielectric effects in magnetic field region of 0–9 T. Mag-netocapacitance effects may also be accompanied with stray contributions that do not necessarily reflect intrinsic magnetoelectric interactions.[25–27] Therefore, observing a magnetic field effect on the ferroelectric polarization loop may be an effective alternative method for studying in depth magnetoelectric coupling. In support of this, the ferroelectric loop at a magnetic field of 11 T [Fig. 3(a)] [region where magnetodielectric effect is big (see Fig. 4] has a different slope near zero electric field compared to the other loops for magnetic fields less than 9 T, where the magnetodielectric effect is small (Fig. 4). It has to be noted that, similarly, magnetic field induced ferroelectric loop has recently been found in Sr substituted $BiFeO_3$ accompanied with no magnetocapacitance effect in this compound.[28]

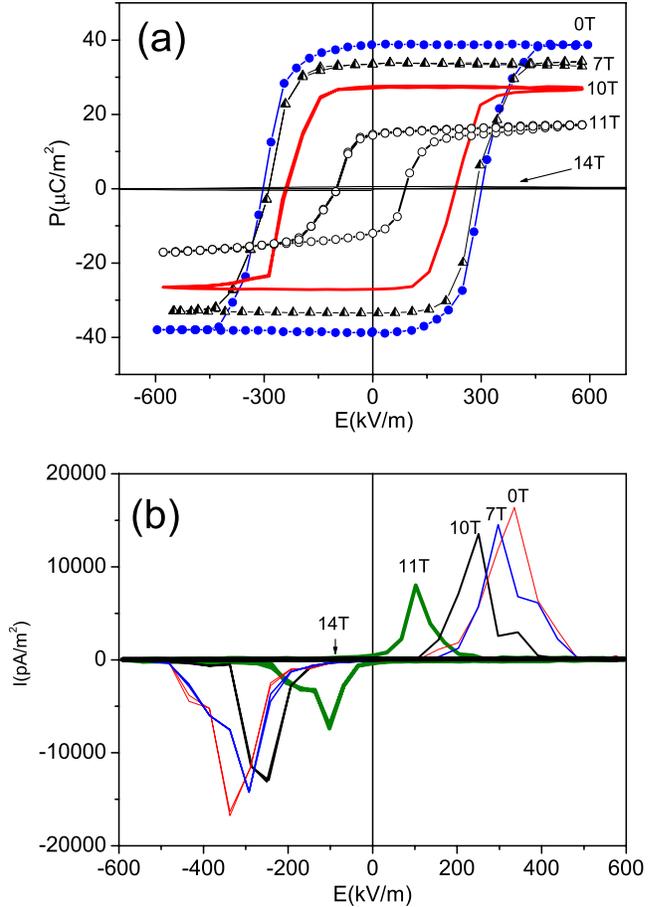

FIG. 3. (Color online) (a) Ferroelectric loops obtained as a result of the ferroelectric current integration at different magnetic fields at 10 K. (b) Voltage-current characteristics taken at 10 K at different magnetic fields.

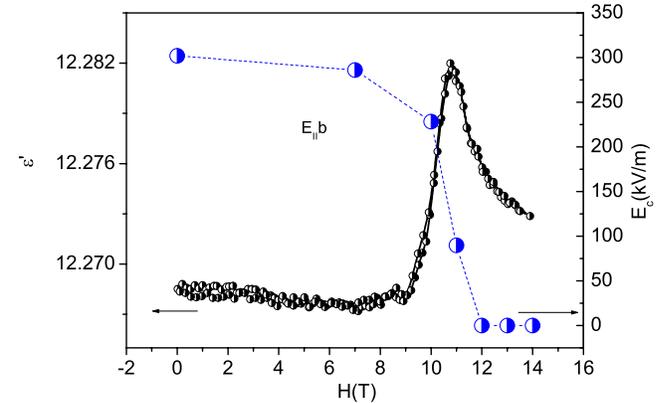

FIG. 4. (Color online) The magnetic field dependence of the dielectric permittivity at 500 kHz (left scale) and ferroelectric coercive force (right scale) at 10 K. The electric and magnetic fields applied parallel to the crystallographic $b$ axis.

In conclusion, a quasistatic technique has been used to investigate ferroelectric properties of a single crystal of $MnWO_4$. It was shown that the sample is indeed ferroelectric and that the shape of its ferroelectric loop strongly depends on both temperature and magnetic field. Increasing the external magnetic field along the b axis decreased both the remanent polarization and ferroelectric coercive force. These effects are not accompanied by any noticeable changes in the magnetic field dependence of dielectric permittivity before the transition to the paraelectric state at about 10.5 T. Therefore, our results also suggest that magnetoelectric coupling may be present without obvious magnetodielectric effects in magnetic and ferroelectric solids.

We thank M. L. Hervé for crystal growth and sample preparation. We also acknowledge the French ANR SEMONE research program.